\documentclass[twocolumn,amsmath,amssymb]{revtex4}
\usepackage{graphicx}
\usepackage{dcolumn}
\usepackage{bm}

\begin{document}

\title{Various representations of the quantity Newton called inertial mass}

\author{J.L. Fry and Z.E. Musielak} 

\affiliation{Department of Physics, The University of Texas 
at Arlington, Arlington, TX 76019, USA}

\begin{abstract}
\noindent
(Received: date / Accepted: date)
\bigskip

\noindent
Newton introduced the concept of mass in his {\it Principia} and gave 
an intuitive explanation for what it meant.  Centuries have passed and 
physicists as well as philosophers still argue over its meaning.  Three 
types of mass are generally identified: inertial mass, active gravitational 
mass and passive gravitational mass.  In addition to the question of what 
role mass plays in dynamical equations and why, the origin of the particular 
amount of matter associated with an elementary particle as a consequence 
of fundamental fields has long been a topic of research and discussion.  In 
this paper, various representations of inertial mass are discussed within the 
framework of fundamental (either Galilean or Poincar\'e invariant) dynamical 
equations of waves and point particles.  It is shown that the derived equations 
have mass-like and mass parameters for waves and point particles, respectively, 
and that the physical meaning of these parameters sheds a new light on the 
fundamental problem of the nature of inertial mass.      
\end{abstract}

\maketitle

\section{Introduction}

The concept of mass was originally introduced by Newton$^{1}$ who wrote in 
his {\it Principia}: 'The quantity of matter is the measure of the same, 
arising from its density and bulk conjointly'.  According to Jammer$^{2}$, 
a major step in interpretation of Newton's concept of mass was made by Euler 
in his {\it Mechanica}$^{3}$, where he suggested that mass should be defined 
as the constant ratio between a constant force and the acceleration caused 
by this force.  Euler's definition of mass had been widely accepted in the 
nineteenth century, however, later in that century, Newton's concept of 
force had become strongly criticized and as a result new definitions of 
mass independent of Newton's second law were proposed.
 
The first step was made by Saint-Venant$^{4}$ in 1845, when he used the 
principle of conservation of linear momentum to express the ratio of masses
of two bodies in terms of their velocity increments after an impact.  Then,
Mach$^{5}$ in 1867 introduced another definition of mass that was based on
two interacting particles, which otherwise were not affected by other 
particles in the Universe. Mach's basic idea was to define mass in terms 
ratios of accelerations caused by the particle's interactions. However, this 
implied the existence of forces whose nature was not specified.  It was also 
pointed out that Mach's assumption of only two particles interacting with 
each other was superficial.  Despite these objections, Mach's definition of 
mass had gained some popularity and become recognized$^{2}$ as "an acceptable 
operational definition of a theoretical construct".

Euler's and Mach's definitions of mass are based on Netwon's second and 
third laws, respectively.  Weyl$^{6}$ also proposed a definition of mass,
which was based on the conservation of momentum.  Both Weyl's and Mach's
definitions have much in common because the conservation of momentum and
the third law have the same physical content, namely, the former is the
time-integrated result of the latter.  A more recent discussion of these
problems can be found in a series of 'Reference Frame' articles written 
for {\it Physics Today} by Wilczek$^{7}$. 
   
There are also other definitions of mass such as that originally introduced
by Hertz$^{8}$ and some based on axiomatized mechanics$^{9}$.  In his definition
of mass, Hertz refers to a number of indestructible and unchangeable particles
at a given point of space and at a given time, and defines mass by weight.  A 
rather different approach is presented in axiomatized mechanics, where Newtonian 
inertial mass can only be determined in Galilean reference frames in which 
the motion of the fixed (very distant) stars must be a disjoint motion$^{2,9}$.        
A different axiomatization of mechanics proposed by Schmidt$^{10}$ intended to
introduce universal concept of mass.  However, the approach was based on the 
existence of Lagrangians, which requires solving the so-called Helmholtz
inverse problem$^{11,12}$.   

Different concepts of mass have also been considered by Pendse$^{13}$, 
Carnap$^{14}$, Kamlah$^{15}$, Zanchini and Barletta$^{16}$, and others.  
A comprehensive review of different concepts of mass can be found in 
Jammer's two books$^{2,17}$, where the second book which was written 
more recently also includes ideas of mass developed in modern physics.  
The book has one chapter devoted to relativistic mass and another dealing 
with the mass-energy relation.  Both concepts have been recently discussed 
by Okun$^{18}$ and Re Fiorentin$^{19}$, who give a new re-interpretation of 
the concept of mass and the relativistic mass-energy relation.

Since the measure of inertia in Special Theory of Relativity (STR) is not 
mass of a particle but its total (kinetic and rest) energy, Okun [18] argues 
that relativistic mass, which depends on particle's velocity, cannot be used 
as the measure of inertia.  He points out that in the very low velocity limit, 
the relativistic rest mass becomes the same as Newtonian mass and therefore
STR and Newtonian mechanics are commensurate theories; see also Jammer's 
discussion in his chapter devoted to relativistic mass$^{17}$.  Now, Re 
Fiorentin$^{19}$ reached similar conclusions, however, his approach was 
different as he used both the Minkowski metric and the principle of least 
action.  His main result that mass is another way of measuring energy requires 
the explanation of the nature of the rest-energy, for which the author refers 
to the Higgs mechanism.
  
The basic idea of the Higgs mechanism is that space is permeated by a scalar 
field, which is called the Higgs field, and that particles couple to this 
field to acquire some energy that can be interpreted as particle's mass$^{20}$.  
More massive particles couple more strongly to the field.  Although this is 
a promissing idea, the scientific community still awaits its experimental 
verification.    

In our previous work$^{21-25}$, we used the Principle of Relativity and the
Principle of Analyticity to formally derive the fundamental equations of
non-relativistic and relativistic mechanics of waves and particles.  For 
the wave mechanics, we considered free and spin-zero elementary particles 
described by scalar wave functions.  We used the extended Galilean group$^{26}$ 
and the Poincar\'e group$^{27}$ to derive the respective Schr\"odinger$^{21}$ 
and Klein-Gordon equations$^{24}$.  We demonstrated that the Schr\"odinger 
equation is the only fundamental (Galilean invariant) dynamical equation 
in Galilean relativity$^{22}$ and that the second-order Klein-Gordon equation 
is the only fundamental (Poincar\'e invariant) equation in space-time with 
the Minkowski metric$^{25}$.  Moreover, we used the same principles to derive 
Newton's equations of non-relativistic and relativistic point particle 
mechanics.  In the derived fundamental equations, we encountered mass-like 
and mass parameters for waves and point particles, respectively.  

The main objective of this paper is to demonstrate the relevance of the
mass-like and mass parameters to the concepts of inertial mass discussed 
above, and to describe various representations of inertial mass within 
the framework of the fundamental (either Galilean or Poincar\'e invariant) 
theories of waves and point particles.  This paper was stimulated by the 
two books on mass written by Jammer$^{2,17}$, and specifically by his 
statement that can be found in the last chapter of the second book:

\bigskip\noindent

"If it were possible to define the mass of a body or particle on its own 
in purely kinematical terms and without any implicit reference to a unit 
of mass, such a definition might be expected to throw some light on the 
nature of mass.  Such a definition, if it existed, would integrate dynamics 
into kinematics and eliminate the dimension M of mass in terms of length 
L and time T." 

\bigskip\noindent

It is now our purpose to show that we have already accomplished the task 
suggested by Jammer, and that our results do indeed shed a new light on 
the fundamental problem of the nature of mass.  

Moreover, we hope that our presentation of the concept of mass will benefit
scientists working in different fields of natural sciences and that it will 
be especially helpful to begining students, who are likely to encounter 
conceptual difficulties with mass in their introductory physics courses.  The 
fact that this indeed can be a serious educational problem was first (to the 
best of our knowledge) recognized by Jackson in his article "Presentation of 
the concept of mass to begining physics students" published in {\it American 
Journal of Physics}$^{28}$ more than 50 years ago.  The author's main point 
is that the concept of mass should be introduced to beginning students by 
discussing various physical phenomena, where mass plays an important role, 
rather than by using formal definitions.  Our paper presents a more modern 
approach to this otherwise old problem.         

The paper is organized as follows.  In Sec. 2, we briefly describe the method 
used to derive invariant dynamical equations in space-time with a given metric, 
and also present the obtained equations.  In Sec. 3, we examine the role of the 
mass-like and mass parameters in the fundamental theories of waves and point 
particles.  In Sec. 4, we determine the relationships between the mass-like and 
mass parameters of the theories, and present various representations of inertial 
mass.  The nature of mass is discussed in Sec. 5, and our conclusions are given 
in Sec. 6.

\section{Fundamental equations of Galilean and Minkowski space-time}

\subsection{Basic procedure}

We are interested in describing a physical object (an elementary particle 
or a classical point particle) by using dynamical equations, which depend 
upon space and time variables that are characterized by a given metric.  
The dynamical equations of a given metric may be derived by the procedure 
used in our earlier work$^{23}$.  Since the procedure explains the appearance 
of mass-like and mass parameters in the derived dynamical equations, we now 
briefly describe it.  The basic procedure of deriving dynamical equations 
for free particles is as follows:

\smallskip\noindent
(i)	Establish a class of observers who define a physical law; for example 
those in isometric frames of reference. 

\smallskip\noindent
(ii) Decide upon the type of theoretical description to be employed; two 
examples are a point particle (classical) description, and a wave description.  
The theory may introduce new quantities, which require an additional metric 
to interpret the dynamical equations, such as the measure of the amplitude 
of a wave in wave theories.

\smallskip\noindent
(iii)	Employ the Principle of Relativity, which states that all observers must 
identify the same physical object and write down the same dynamical equations 
describing its space-time evolution.  This could equally well be taken as the 
definition of a law instead of a principle.  Clearly changing the class of 
observers could change the form and apparent nature of the laws.

\smallskip\noindent
(iv)	Employ the Principle of Analyticity, which requires that all things 
that can be measured must be described by analytic functions of the space-time 
variables.

\subsection{Galilean and Poincar\'e invariant equations}

In our previous work, we considered free and spin-zero elementary particles 
described by scalar wave functions.  To derive Schr\"odinger and Klein-Gordon 
equations, we used the extended Galilean group$^{26}$ and the Poincar\'e 
group$^{27}$, respectively, and obtained 
\begin{equation}
\left [ i {{\partial} \over {\partial t}} + {1 \over {2M}}
\nabla^2 \right ] \psi = 0\ ,
\label{IIeq1}
\end{equation}
and
\begin{equation}
\left [ \partial^{\mu} \partial_{\mu} + {{\omega_0^2} \over {c^2}} 
\right ] \phi = 0\ ,
\label{IIeq2}
\end{equation}

\noindent
where $\psi$ and $\phi$ are scalar wave functions, $M$ and $\omega_0$ 
are the so-called wave mass$^{21}$ and invariant frequency$^{23,24}$, 
respectively, and $c$ is the speed of light.   

In addition to free and spin-zero elementary particles described by scalar 
wave functions, we also considered free classical point particles$^{29}$ and 
derived both non-relativistic and relativistic versions of Newton's second 
law of dynamics.  The obtained equations can be written in the following 
form:
\begin{equation}
m {{dV^i} \over {dt}} = 0\ , 
\label{IIeq3}
\end{equation}
and
\begin{equation}
M_0 {{dU^{\mu}} \over {d \tau}} = 0\ , 
\label{IIeq4}
\end{equation}

\noindent
where $\tau$ is the proper time, $V^i$ is the three-velocity vector with 
$i = 1$, $2$ and $3$, and $U^{\mu}$ is the four-velocity vector with 
$\mu = 0$, $1$, $2$ and $3$.  In addition, $m$ represents Newtonian 
inertial mass that is measured in $kg$ in the SI system of units, and 
$M_0$ is a derived parameter whose units are chosen here to be the same 
as the wave mass $M$. 
  
In deriving the above dynamical equations, we encountered the need for 
the four parameters ($M$, $\omega_0$, $m$ and $M_0$) that describe the 
elementary particles and have the same value in all inertial frames of 
reference.  Each of these parameters is a manifestation of inertial mass 
of an elementary particle, so we call $M$ and $\omega_0$ the mass-like
parameters, and $m$ and $M_0$ the mass parameters; we call $M_0$ the 
mass parameter despite its units, which are the same as $M$, because 
it represents mass of a point particle.  Examining the invariant 
dynamical equations for free particles, we can offer an interpretation 
for the meaning of each of the invariant constants describing a free 
particle in the above four different dynamical equations.   

\section{Invariant mass-like and mass parameters} 

In the Galilean metric, the Schr\"odinger equation given by Eq. (\ref{IIeq1}) 
contains one single parameter $M$, which is Galilean invariant and we call 
it the wave mass$^{21}$.  The origin of this parameter is the definition of 
an elementary particle. The wave vector $\vec k$ and the frequency $\omega$ 
are the eigen-labels by which its wave representation may be labeled in free 
space, and the Galilean invariant ratio of these labels upon which all inertial 
observers must agree$^{21,23}$: $M = k^2 / 2 \omega$.  Now, $M$ may be determined 
independently of (Newtonian) mass $m$, has units derived from space and time only, 
and is listed for various elementary particles in Table 2 of our paper$^{23}$.  
It occurs naturally and cannot be avoided in a Galilean wave description of 
an elementary particle.  

From the dispersion relation $\omega / k^2 = 1 / 2 M$, we deduce that 
if a particle is caused to change its state to a new value of $k$ in 
a given frame of reference, then the change in $\omega$ is proportional 
to $1/M$.  The larger $M$, the smaller the change in the state label 
$\omega$.  Thus $M$ measures the resistance to change in frequency of 
the state of a free particle, a property we relate to the inertia of 
the particle.

In the Minkowski metric, the Klein-Gordon equation (see Eq. \ref{IIeq2}) 
contains a single parameter $\omega_0$, which is Poincar\'e invariant 
and we called it the invariant frequency in our previous paper$^{24}$.  
The origin of this parameter is the requirement of a Poincar\'e invariant 
description of an irreducible representation of the Poincar\'e group.  
While $\vec k$ and $\omega$ must also be eigen-labels of the irreducible 
representations (irreps) of the Poincar\'e group in any inertial frame 
of reference, a Poincar\'e invariant label is the length of the eigen 
four-vector $k^{\mu}$, where $k^{0} = \omega/c$.  The invariant frequency 
may be determined independently of wave mass and Newtonian mass but it is 
related to them$^{24}$.  Its units are a derived quantity, depending upon 
units of time only.  

Values of invariant frequencies for various particles are listed in 
Table 2 of our paper$^{24}$.  The parameter $\omega_0$ is a measure of 
the inertial properties of matter, occurs naturally and cannot be 
avoided in a Poincar\'e wave description of an elementary particle.  
In a given frame of reference the dispersion relation $\omega^2 = 
\omega_0^2 + k^2$ allows us to deduce that the greater $\omega_0$, 
the smaller the change in $\omega$ for a given change in $k$.  
Thus $\omega_0$ is a measure of a particle's resistance to change 
in frequency $\omega$ of the state of the elementary particle in 
a given frame of reference, a property we relate to the inertia 
of the particle. 

\begin{table}
{\offinterlineskip \tabskip=0pt
\halign{ \strut \vrule#& \quad \bf# \quad  &
\vrule#&
\quad
\hfil # \quad &
\vrule#&
\quad
\hfil # \quad &
\vrule#&
\quad
\hfil # \quad &
\vrule#&
\quad
\hfil # \quad &
\vrule#&
\quad
\quad
\hfil # \quad &
\vrule#
\cr
\noalign{\hrule}
\noalign{\hrule}
&       &&              &&           &\cr
& Metric/Theory && Invariant && Dispersion &\cr
&            && parameter && relation &\cr
\noalign{\hrule}
&       &&              &&           &\cr

& Galilean/wave && $M$ && $k^2 = 2 M \omega$ &\cr

&       &&              &&           &\cr

& Minkowski/wave && $\omega_0$ && $\omega^2 - k^2 = \omega_0^2$ &\cr

&       &&              &&           &\cr

& Galilean/particle && $m$ && $p^2 = 2 M E$ &\cr

&       &&              &&           &\cr

& Minkowski/particle && $M_0$ && $P^{\mu} P_{\mu} = M_0^2$ &\cr

&       &&              &&           &\cr
\noalign{\hrule} }}
\bigskip
\bigskip
\caption{Invariant parameters and dispersion relations for the wave
and point particle theories in space-time with the Galilean and 
Poincar\'e metrics.}
\end{table}

The form of the free particle dynamical equations in point particle 
theories is very different from that of the wave equations.  The 
parameters remaining after setting the forces equal to zero on the 
RHS of Eqs (\ref{IIeq3}) and (\ref{IIeq4}) are $m$ and $M_0$, 
respectively.  The parameter $m$ in Newtonian mechanics is customarily 
assigned a new fundamental unit of measure, the $kg$ in the SI system 
of units, while $M_0$ is a derived parameter which we have chosen to 
have the same units as wave mass $M$.  As already shown by us$^{24}$,
$M_0$ may be related to $m$ as well as to $M$ and $\omega_0$. 

The invariant mass-like and mass parameters for the wave and point 
particle theories given by Eqs (\ref{IIeq1}) through (\ref{IIeq4})
are listed in Table 1, which also contains the corresponding 
dispersion relations.  Other local parameters for these theories, 
the three-vectors $k^i$, $p^i$ and $P^i$, and the scalars $\omega$, 
$E$ and $P^0$ are given in Table 2, with $p^i = m v^i$ and $P^{\mu} 
= M_0 dx^{\mu} / d\tau$.  The three-vectors and scalars are also 
acceptable labels in a given frame of reference, however, they 
differ in value from one inertial frame of reference to another. 
  
\begin{table}
{\offinterlineskip \tabskip=0pt
\halign{ \strut \vrule#& \quad \bf# \quad  &
\vrule#&
\quad
\hfil # \quad &
\vrule#&
\quad
\hfil # \quad &
\vrule#&
\quad
\hfil # \quad &
\vrule#&
\quad
\hfil # \quad &
\vrule#&
\quad
\quad
\hfil # \quad &
\vrule#
\cr
\noalign{\hrule}
\noalign{\hrule}
&       &&              &&           &\cr
& Metric/Theory && Scalars && Vectors &\cr
\noalign{\hrule}
&       &&              &&           &\cr

& Galilean/wave && $\omega$ && $k_i$ &\cr

&       &&              &&           &\cr

& Minkowski/wave && $\omega$ && $k_i$ &\cr

&       &&              &&           &\cr

& Galilean/particle && $E$ && $p$ &\cr

&       &&              &&           &\cr

& Minkowski/particle && $P_0$ && $P_i$ &\cr

&       &&              &&           &\cr
\noalign{\hrule} }}
\bigskip
\bigskip
\caption{Frame of reference dependent labels for the wave and point 
particle theories in space-time with the Galilean and Poincar\'e 
metrics.}
\end{table}

Since elementary particles in Nature appear to be best described 
by wave equations, which have parameters with derived units, the 
description of inertial mass by an additional fundamental measure, 
the kilogram, is possible but unnecessary.  For elementary particles 
it is less accurately known than the corresponding wave mass$^{23,24}$ 
and thus it should not be the measure of choice.  The dynamical 
equations for free classical point particles given by Eqs (\ref{IIeq3}) 
and (\ref{IIeq4}) have solutions independent of the mass parameters 
$m$ or $M_0$; the trajectories and world lines are the same for all 
values of these parameters.    

\section{Relationship between parameters of theories}

\subsection{Galilean metric}

Consider a free particle moving in space-time with a Galilean metric 
and characterized by $k_i$, $\omega$ and $M$ in the wave description 
and by $p_i$, $E$ and $m$ in the point particle description.  Let us 
assume that it is possible to arrange an interaction with a field so 
that both wave and particle descriptions may be employed in determining 
the parameters associated with the elementary particle.  These conditions 
are described in most derivations of Ehrenfest's theorem and we assume 
they can be achieved for purpose of discussion here.  Using the free 
particle parameters listed above, one observer determines the direction 
of travel of a wave and the other determines the direction of travel of 
what he assumes to be a point particle.  Since it is in fact the same 
object their local vector parameters $\vec k$ and $\vec p$ must be 
parallel, so their magnitudes differ only by a real constant.  Thus, 
we may write $\lambda k_i = p_i$ and obtain from the dispersion 
relations
\begin{equation}
E = \lambda^2 \omega {M \over m}\ .
\label{IVeq1}
\end{equation}

Here $\lambda$ is an arbitrary real constant.  We note that from its 
definition the units of classical mass are arbitrary, i. e. changing 
them changes the units of force and energy but not trajectories computed 
from Newton's second law.  On the other hand the units of wave mass are 
established from the choice of units of length and time.   We may choose 
to measure $m$ and $M$ in the same units so that for the same particle 
they are equal.  Then the units of $E$ are the same as the units of 
$\omega$ if we $\lambda = 1$, a dimensionless number.  

On the other hand it is customary to interpret Eq. (\ref{IVeq1}) by 
writing $\lambda M = m$ and $E = \lambda \omega$ using the experimentally 
determined value of $\lambda$, which is of course known as the Planck 
constant, $\hbar$.  The wave equations given by Eqs (\ref{IIeq1}) and 
(\ref{IIeq2}) were both derived without any reference to the Plank 
constant and contain only the parameters $M$ and $\omega_0$, both of 
which can be determined without reference to the Planck constant.  
Since the usual classical mass introduces an unnecessary fundamental 
unit into physics, we prefer relating $m$ to the wave mass $M$ for the 
same particle.  For elementary particles the wave mass may be determined 
to almost two orders of magnitude less residual error than the residual 
errors in the Planck constant or classical mass.  Thus the wave equations 
without classical mass and the Planck constant are more accurate and they 
should be used to describe elementary particles$^{23}$.

\subsection{Minkowski metric}

The relativistic wave equation and the relativistic point particle equation 
are completely independent of each other, but in an appropriate limit the 
relativistic wave may be interpreted as a point particle$^{24,25}$.  Using 
the results of these papers, we relate $k^{\mu}$ and $P^{\mu}$ to $\omega_0$ 
and $M_0$ by using $k^{\mu} k_{\mu} = \omega_0^2$ and $P^{\mu} P_{\mu} = M_0^2$.  
Since $k^{\mu}$ and $P^{\mu}$ both provide the direction of motion along the 
world line for the same particle under the proper experimental setup so that 
both theories are valid, the two four-vectors are parallel and can differ 
only by their lengths.  Since $M_0$ has arbitrary units, its units can be 
chosen so that the lengths are the same: $\omega_0=M_0$.  

In general, we have $\omega_0 = M_0 c^2$ and $M_0$ has the same units as wave 
mass $M$, a unit derived from $L$ and $T$.  However, if the units with $c = 1$ 
are used, then both $\omega_0$ and $M_0$ may be expressed in units of $1/T$.  
Because of this relationship between the invariant frequency $\omega_0$ and 
the rest mass $M_0$ it is possible to remove the fundamental definition of 
relativistic mass and replace it with a derived unit of mass as it was already 
done in Galilean relativity (see the previous subsection).  We note that the 
Planck constant did not enter in this relationship.  The dynamical equation of 
point particles in the Minkowski metric may be expressed in terms of wave mass 
units.

The non-relativistic limits of Eqs (\ref{IIeq2}) and (\ref{IIeq4}), given in 
some textbooks$^{30,31}$, lead to additional connections between the mass-like 
and mass parameters of the two metrics.  Thus with units $c = 1$, we obtain 
$m = M_0$ and $M = \omega_0$ when $m$ and $M$ are measured in units of $1/T$ 
instead of kilogram units.  Combining all relationships between the invariant 
parameters, an elementary particle in a Minkowski metric may be described under 
appropriate conditions by any one of four dynamical equations with all invariant 
mass-like parameters being the same:
\begin{equation}
m = M = \omega_0 = M_0\ .
\label{IVeq2}
\end{equation}

In this process the familiar concepts of mass, length and time, which are 
considered fundamental units of Nature, have been replaced by one fundamental 
unit for time, and mass and length units have been reduced to derived units 
$1/T$ for mass and $T$ for length.  Thus there is no need for a circular 
definition of mass and the units of space and time are properly connected 
in the Minkowski metric.  The unit of mass was eliminated by the connection
to the wave equations and the unit of space was eliminated by the Minkowski
metric.  The wave equations appear to have eliminated the circular definition
of mass critized by Jammer$^{2,17}$. 

\subsection{Various representations of Newton's inertial mass}

According to the above results, Netwon's inertial mass may be represented 
by different mass-like and mass parameters that arise in the fundamental 
(Galilean or Poincar\'e invariant) equations of waves and point particles.  
To obtain this important result, we assumed that the most basic elements of 
our approach were the metrics, which we used to define elementary particles, 
derive the invariant dynamical equations, and determine the corresponding 
mass-like and mass parameters.  

By studying the relationships between Newton's inertial mass and these 
parameters, we established that all inertial observers must agree upon 
the value of the mass in order to identify the same elementary particle.  
Dynamical behavior of free elementary particles is governed by the 
mass-like and mass parameters and by the way they enter each invariant 
dynamical equation.  Their presence in the Galilean and Poincar\'e 
invariant dynamical equations leads to properties that we identify 
physically with Newton's original concept of inertia. 

In Newtonian mechanics, the property identified as inertia is commonly 
known as a resistance to a change in velocity of a particle with mass 
$m$, which is called the inertial mass.  A generalization of this 
property, valid for all four fundamental theories considered above, 
is that the mass-like and mass parameters reflect the resistance of a 
particle to a change in its free particle state.  The principal effect 
of a larger mass-like (or mass) parameter is to make it more difficult 
to increase the energy-like measures of the state of the system as the 
momentum-like parameters are increased upon application of a given force.  
This concept has been used to provide a working definition of a classical 
elementary particle$^{29}$.  

\section{The nature of mass}

Mass occurs naturally in our invariant dynamical equations as a result of 
type of metric, definition of physical law, definition of an elementary 
particle, assumption of analyticity, and resulting differential equations.  
The central idea is that mass labels the irreps of the group of the metric, 
and that it also characterizes the nature of the state function during its 
transformation from one isometric frame of reference to another.  Thus, in 
our approach, mass is a natural consequence of the Galilean and Minkowski 
metrics.  

Some understanding of the inertial properties of mass can be gained from 
the work of Barut$^{32,33}$, who demonstrated that it is possible to take 
wave equations for massless particles and by separating variables find 
a localized solution corresponding to a rest frame frequency $\omega_0$.  
The equations then appear to have properties of a wave equation with mass 
proportional to the invariant frequency $\omega_0$.  Based on the results 
presented in this paper, as well as on Barut's results, we conclude (Barut 
did not state so) that localization is the process by which inertial mass 
appears.  What causes the localization with observed elementary particle 
frequencies is not fully understood for all particles, but interesting 
accounts of most the neutron and proton inertial masses have been given 
by Wilczek$^{34}$ in his "What Matters for Matter" discussions presented 
in {\it Physics Today}. 

We have accomplished the task suggested by Jammer (see Sec. 1) by defining 
the mass of an elementary particle on its own without any specific reference 
to the unit of mass 'kilogram'.  This elimination of the dimension of mass 
has allowed us to formulate the fundamental quantum theories based on the 
Schr\"odinger and Klein-Gordon equations without making any reference to 
the Planck constant.  We have also contributed to the challenging problem 
of the nature of mass by showing that the mass-like and mass parameters are 
related to the concept of inertial mass originally introduced by Newton, and
that among these parameters the invariant frequency $\omega_0$ is the most 
fundamental one as the other parameters may be derived from it.  Why only 
selected values of this parameter occur in Nature must be determined from 
considerations other than the free particle dynamical equations considered 
here.  Why $\omega_0$ takes the special values observed in Nature is not 
fully understood for elementary particles, but thought to arise from some 
underlying fields, like for instance the Higgs field$^{20}$.  

\section{Conclusions}

We have discussed the fundamental dynamical equations for waves and 
point particles in space-time with both the Galilean and Minkowski 
metrics.  The obtained equations are either Galilean or Poincar\'e 
invariant and they describe free spin-zero elementary particles that 
are represented by scalar wave functions, and free classical particles 
that are treated here as point particles.  There are four invariant 
mass-like and mass parameters in these equations, and we have shown 
that these parameters are various representations of Newton's inertial 
mass.  Our discussion of the relationships between the parameters and 
their physical meaning sheds a new light on the fundamental problem of 
the nature of inertial mass. 
   
From the perspective of this paper, inertial mass is just a parameter 
that all inertial observers must agree upon to identify elementary 
particles.  The particular way in which the inertial mass-like and 
mass parameters enter each invariant dynamical equation governs its 
dynamical behavior, leading to properties that we identify physically 
with the concept of inertia.  Inertial mass is a frame of reference 
independent description of the particle, while energy-like and 
momentum-like labels on the free particle are frame of reference 
dependent.  The latter two quantities are nonetheless very useful 
in the description of the state of a particle relative to a given 
frame of reference. 

It is our hope that our presentation of the concept of mass given in 
this paper will be helpful to scientists working in different fields 
of natural sciences and that it will especially benefit begining 
students, who are likely to encounter conceptual difficulties with 
mass in their introductory physics courses.  The main message of this
paper for the begining students is that the concept of mass occurs in 
Nature naturally once the metric of space-time in which we live is 
determined.  Moreover, the mass of an elementary particle can be 
defined on its own, without any reference to the specific unit of
mass 'kilogram'.  This has important physical consequences as it 
allows formulating the fundamental quantum theories without formally 
introducing the Planck constant but using what is called here a mass-like 
parameter.  Hence, the theories of physics may be formulated by using 
either the classical concept of mass with its unit of 'kilogram' and 
the Planck constant, or by using only one the mass-like parameter.  
It must also be pointed out that the theories based on the mass-like
parameter can attain higher accuracy of performing computations$^{23,24}$.       

\bigskip\noindent
{\bf ACKNOWLEDGMENTS}
We thank Alex Weiss and Andy White for discussions and their 
comments on the manuscript.  Z.E.M. acknowledges the support of 
this work by the Alexander von Humboldt Foundation.  
%


\end{document}